\documentstyle[aps,preprint]{revtex}
\begin{document}
\draft

\title{Peak Effect in Superconductors: Melting of Larkin Domains}

\author{Xinsheng Ling,$^{(1)}$ Chao Tang,$^{(1)}$ S. Bhattacharya,$^{(1)}$
 and Paul M. Chaikin$^{(1,2)}$}
\address{$^{(1)}$NEC Research Institute, 4 Independence Way, Princeton,
NJ 08540}
\address{$^{(2)}$Department of Physics, Princeton University, Princeton,
NJ 08544}

\date{April 25, 1995}

\maketitle

\begin{abstract}

Motivated by the recent observations of the peak effect in high-$T_c$ YBCO
superconductors, we reexamine the origin of this unusual phenomenon.  We
show that the mechanism based on the $k$-dependence (nonlocality) of the
vortex-lattice tilt modulus $C_{44}({\bf k})$ cannot account for the essential
feature of the peak effect.  We propose a scenario in which the peak effect
is related to the melting of Larkin domains.  In our model, the rise of
critical current with increasing temperature is a result of a crossover
from the Larkin pinning length to the length scale set by thermally excited
free dislocations.

\end{abstract}


\newpage
\narrowtext

About 35 years ago Le Blanc and Little \cite{blanc} discovered a striking
phenomenon, later known as the ``peak effect'', in a conventional
superconductor Nb that the sample can carry more supercurrent at a higher
temperature (or field) slightly below $B_{c2}(T)$ where it becomes normal.
Over the years, the peak effect was found to be ubiquitous in conventional
superconductors \cite{peak,peak2d,peak-shobo} and it has been observed
recently in the high-$T_c$ superconductor YBCO crystals \cite{ling,kwok}.
Pippard \cite{pippard} and Larkin and Ovchinnikov \cite{lo} proposed that
the peak effect is a result of an anomalous softening of the vortex lattice.
The basic physics is that a soft lattice can be pinned more strongly than
a more rigid lattice.  In fact, an infinitely rigid lattice cannot be
pinned at all by random pinning.  The unresolved problem, however, is the
mechanism which leads to the abrupt loss of the vortex-lattice rigidity.

In this paper, we first reexamine the standard interpretations of the peak
effect and show that the mechanism of an anomalous softening of the
wavevector dependent tilt modulus $C_{44}({\bf k})$ does not account for the
essential features of the peak effect.  We then propose a scenario in
which a melting of the ``Larkin domains'' leads to the peak effect.  The
rise of the critical current with increasing temperature is a result of
a crossover of two length scales: from the Larkin pinning length to the
average separation between thermally excited free dislocations
\cite{lt}.

Let us first recall briefly the general features of the peak effect
phenomenon in both conventional superconductors and high-$T_c$ superconducting
YBCO single crystals.  In samples with high values of the critical current
density $j_c$ at a fixed magnetic field $B$, one usually finds that $j_c$
decreases to zero monotonically with increasing temperature and the peak
effect is absent \cite{peak,peak-shobo,ling,kwok}.  But in samples with low
$j_c$ (weak pinning), the temperature dependence of $j_c$ can be quite
different.  Fig.\ \ref{data} is a plot of the critical current density
as a function
of temperature extracted from Ref.\ \cite{ling} for a YBCO crystal.  With
increasing temperature, $j_c$ initially decreases monotonically, then
suddenly rises, reaches a peak before finally dropping to zero.
Experimentally, the peak effect is identified as a dip in resistance
\cite{peak,peak2d,peak-shobo,kwok}, a dip in the in-phase part of ac
susceptibility \cite{ling}, or a peak in critical current density $j_c$
(obtained with the standard voltage criterion)
\cite{peak,peak2d,peak-shobo,kwok} as a function of $T$ or $B$.  In Nb
and other low-$T_c$
superconductors, the onset (the rise of $j_c$) temperature $T_p$ of the
peak effect is very close to $T_c(B)$ and $T_c(B)-T_p \sim$ 0.5 K.  In
YBCO, $T_c(B)-T_p \sim$ 5 K is about 10 times larger.  However,
$(T_c(B)-T_p)/T_c(B) \sim$ 0.05--0.1 is about the same for both low-$T_c$
and high-$T_c$ superconductors.  The problem under consideration here is
why the critical current rises with increasing temperature.

Pippard \cite{pippard} proposed that the rise of the critical current has
to result from a rapidly decreasing rigidity of the vortex lattice.  The
rigidity of the vortex lattice prevents the vortex lines from taking
advantage of the valleys of random pinning potential.  Thus a rapidly
decreasing rigidity would allow the lattice to conform better to the
pinning potential and enhance critical current.  This idea on pinning was
subsequently put forward more rigorously by Larkin and Ovchinnikov (LO)
in their theory of collective pinning \cite{lo}.  It was shown by Larkin
\cite{larkin} that in the presence of random pinning the vortex lattice
loses its long-range translational order and breaks up into domains of
correlated regions in which the vortex lines interact elastically.  LO
argued \cite{lo} that the critical current density is determined by the
fluctuations of random potential in a domain and the pinning force
density $j_cB = (nf^2/V_c)^{1/2}$, where $n$ and $f$ are the density
and strength of the pins, $V_c$ the volume of the domain.  The size of
the Larkin domains can be estimated by a simple energy consideration.  The
vortex lattice deforms to take advantage of the random pinning potential at
the cost of the elastic energy.  The total unit volume energy change is
\cite{lo}
\begin{equation}
\delta F = C_{66} (\frac{r_p}{R})^2 + C_{44} (\frac{r_p}{L})^2
            - fr_p(\frac{n}{V})^{1/2},
\label{energy}
\end{equation}
where $C_{66}$ is the shear modulus of the lattice, $r_p$ the range of
the pinning potential, $R$ and $L$ are the transverse (to the field)
and longitudinal (along the field) dimensions of the domain, and $V = R^2L$.
The minimization of Eq.~(\ref{energy}) gives the pinning lengths $R_c$
and $L_c$: $R_c \sim C_{66}^{3/2}C_{44}^{1/2}r_p^2/nf^2$, $L_c =
(C_{44}/C_{66})^{1/2}R_c$.  In very thin samples with a perpendicular
field, if the pinning is so weak that $L_c$ is greater than the sample
thickness, the problem becomes two-dimensional (2D) and only $R_c \sim
C_{66}r_p/n^{1/2}f$ is relevant.  In the LO theory, since $j_cB =
(nf^2/V_c)^{1/2}$, the peak effect can be accounted for if the volume of
Larkin domain $V_c$ drops faster than $nf^2$ in some field or temperature
range.  The central question here is what mechanism does that.

It was found by Brandt \cite{brandt1} that, near upper critical field
$B_{c2}$ the vortex fields overlap strongly and the tilt modulus $C_{44}$
becomes nonlocal: it softens substantially for short wavelength tilt
deformation.  Most experiments of the peak effect on low-$T_c$ materials
are carried out as a function of field while keeping the temperature
constant, the peak effect manifests itself as a peak in $j_c$ (or a dip
in resistance) near $B_{c2}(T)$.  It is thus natural to relate the
$C_{44}({\bf k})$ softening mechanism to the peak effect.  LO found
\cite{lo} that this mechanism leads to an exponential form for
$V_c$ when $R_c$ becomes smaller than $\lambda^\prime =\lambda
/(1-b)^{1/2}$, where $\lambda$ is the penetration depth and $b= B/B_{c2}$,
and $j_c\sim \exp(-BC_{66}^{3/2}k_hr_p^2/W)$, with $W = nf^2$
and $k_h = 1/\lambda^\prime$.  It was customary to assume \cite{lo} a
scaling function for the field dependence of $W$, $W \sim b(1-b)^2$,
and with $C_{66} \sim b(1-b)^2$ \cite{brandt2} and $k_h^2 \sim (1-b)$,
one indeed finds that $j_c$ rises exponentially with $B$ for $b \sim 1$.

The above interpretation of the peak effect has two major difficulties.
The first is that if this mechanism is the relevant one, it should also
account for the temperature dependence: $j_c$ rises with increasing $T$
in the peak regime.  Giving the most liberal estimate for the temperature
dependence of the parameters, however, the above mechanism fails to explain
why $j_c$ rises with $T$.  The elementary pinning force $f$ is a function
of the local gradient of the amplitude of the gap function, or $f \sim
a_1|\Delta(r)|^2$.  For $T_c$ smearing pins, $a_1 \sim (1-t)$,
$|\Delta(r)|^2 \sim (1-t$), $t = T/T_c$, $f \sim (1 - t)^2$, and
for pins that do not smear $T_c$, $f \sim (1-t)$.  Thus $W \sim (1-t)^2$
or $(1-t)^4$.  Without melting, $C_{66} \sim (1 - t)$ and $k_h \sim
(1-t)^{1/2}$.  Thus $j_c$ either does not change with $T$ or decreases
exponentially with increasing temperature and $j_c$ never increases with
increasing $T$, according to this mechanism.  The second difficulty of
this mechanism is that the peak effect has been observed in thin films
\cite{peak2d} and in very thin NbSe$_2$ crystals with pinning weak enough
such that $L_c$ exceeds the sample thickness \cite{peak-shobo,koo}, in which
$C_{44}$ does not seem to play any role.  Therefore we believe that
this mechanism cannot be responsible for the peak effect.

Another modulus of the vortex lattice which enters the pinning problem is
the shear modulus $C_{66}$.  The peak effect, the rise of $j_c$, suggests
to us that $C_{66}$ of the vortex lattice vanishes at the peak effect
regime much faster than what was calculated \cite{brandt2} from the
Ginzburg-Landau theory.  In fact, Pippard \cite{pippard} suggested that
the peak effect could be due to the softening of the shear modulus of the
vortex lattice.  Here we propose that {\it the rise of the critical
current, or the peak effect itself, is a result of vortex lattice melting.}

Indeed, the fact that a vortex lattice can melt has been pointed out by
several authors \cite{huber,fisher,nelson}.  Most of the efforts has been
focused on the possibility of melting of a perfect Abrikosov lattice
and the determination of the phase diagrams.  A 2D vortex lattice can melt
via the mechanism of the unbinding of the dislocation pairs
\cite{huber,fisher}, similar to the melting of a 2D crystal studied by
Kosterlitz and Thouless \cite{kt}, and by Halperin and Nelson \cite{hn}.
Much less is known for the melting of a 3D lattice in general.  For a
perfect 3D vortex lattice both analytic considerations \cite{brezin} and
numerical simulations \cite{hetzel} suggest a first order transition.  In
particular, it has been shown that a finite density of free edge
dislocations would result in a zero long wavelength shear modulus
\cite{marchetti}.

Generally speaking, the melting of a lattice (2D and 3D) in the presence
of quenched disorder is not well understood.  Indeed, often it is not clear
what ``melting'' means in this situation.  Naively, if the pins are dilute
compared to the vortex density, the melting would be more or less the same
as in a pure system and would result in a free (unpinned) vortex fluid.
It is doubtful that such a regime exists in real samples.
In fact, almost all weak-pinning (low $j_c$) samples show peak effect
\cite{peak,peak-shobo,ling}, which implies a high density of pins
according to our scenario described below.  It was argued \cite{ffh}
that a vortex lattice, disordered due to random pinning,
could be a ``vortex glass'' at low temperatures, and the melting
transition is then a vortex-glass-to-vortex-fluid transition, which is
second order at least for strong random pinning.  While a vortex glass
phase, if it exists, would determine the small current behavior especially
for systems with strong disorder, the peak effect is a property of
(relatively) large current in samples with weak pinning.

For a vortex lattice in the presence of random pinning, the lattice
deforms locally to take advantage of the fluctuations of the random pins
and the long range order of the lattice is destroyed \cite{larkin}.  The
lattice is ordered within Larkin domains whose size is the pinning lengths
$R_c$ and $L_c$.  (We assume weak pinning by which we mean that the
typical pinning force $f$ for individual pins is small.  We shall always
assume that the density of pins $n$ is not small compared with the vortex
density, which seems to be the case even in samples with extremely low
$j_c$.)  Note that $R_c$ and $L_c$ are also
the length scales beyond which the lattice is elastically decoupled.  In
other words, a small local shear deformation of the lattice would not
propagate (elastically) much farther than $R_c$.  How would such a weakly
pinned lattice melt?  Melting of a lattice is usually characterized by the
vanishing of the long wavelength shear modulus.  For a pinned vortex
lattice, however, a long wavelength shear is not sustained by the long
wavelength elasticity of the lattice.  If we shear slightly the opposite
sides of a pinned vortex lattice, the shear deformation would decay
(presumably exponentially) inside the lattice with a decay length of order
$R_c$, and the vortices inside would not feel the shear.  In other words, the
long wavelength shear is sustained by the pinning force.  It is then not
surprising that the length scale $R_c$ should play a crucial role in the
melting of a pinned lattice.

We first consider the case of a 2D lattice where much is known for the
melting of a clean (pinning free) system \cite{kt,hn}.  In the case of a
2D lattice, thermally excited dislocation pairs are bound for temperatures
below the melting temperature $T_m$ and, consequently, the shear modulus is
finite.  At $T_m$, the largest dislocation pairs start to dissociate and the
long wavelength shear modulus drops discontinuously to zero.  Above $T_m$,
the density of free dislocations rises from zero, and the mean distance
between free dislocations is the Kosterlitz-Thouless correlation length
\cite{foot-kt} $\zeta \sim \exp[c/(T-T_m)^{\bar\nu}]$ with $\bar\nu\sim
0.37$ \cite{hn}, which diverges as $T\rightarrow T_m^+$.  The correlation
length $\zeta$ also sets the length scale for the $q$(wavevector)-dependence
of the shear modulus $\mu(q,T)$: roughly speaking, $\mu$ is zero for
$q<1/\zeta$ and finite for $q>1/\zeta$.  Now imagine that the vortex
lattice is weakly pinned.  For $T<T_m$, the Larkin length $R_c$ sets the
elastic length scale and the critical current density $j_c=n^{1/2}f/R_cB$.
At $T=T_m$, the vortex lattice melts with the long wavelength shear
modulus dropping to zero.  However, as argued in the previous paragraph,
the pinned lattice would not feel being melted at this point since the
lattice is elastically decoupled beyond the length scale of $R_c$.  The
critical current density is still determined by $R_c$.  For $T>T_m$,
another length scale $\zeta$, the Kosterlitz-Thouless correlation length,
enters the system.  $\zeta$ decreases exponentially fast from the
infinity as the temperature is increased and will soon become comparable
to $R_c$.  For $\zeta < R_c$, the Larkin domains melt and the relevant
elastic length scale for the determination of $j_c$ is now $\zeta$:
\begin{equation}
j_c = \frac{n^{1/2}f}{\zeta B} \approx
      \frac{n^{1/2}f}{B}\exp[-\frac{c}{(T-Tm)^{\bar\nu}}], \quad\quad
       (\zeta < R_c).
\label{j-2d}
\end{equation}
{\it Thus the onset the of peak effect occurs when the two length
scales $R_c$ and $\zeta$ cross each other} (Fig.\ \ref{sketch}).
The exponential increase of $j_c$ with $T$ (Eq.~(\ref{j-2d})) would
continue until $\zeta$ is of the order of the lattice constant. However,
the temperature dependence of the pinning force $f$ and thermally
activated vortex motion \cite{foot-an} would presumably dominate the
behavior of $j_c$ at even higher $T$ and cause $j_c$ to vanish.

In the 3D case, melting of a pure lattice is much less understood.  If the
melting transition is mediated by generation of the free edge dislocations,
one would expect a similar mechanism for the peak effect as
in 2D, with $\zeta$ now being the mean distance between dislocation lines.
To make a qualitative or semi-quantitative estimate for $j_c$ in the peak
effect regime, we take the Landau-Ginzburg-like free energy often used
in 3D dislocation systems \cite{yama}:
\begin{equation}
F(\rho) = - F_1 \rho \ln C\rho + F_2 \rho + F_3 \rho^2,
\label{disloc}
\end{equation}
where $\rho$ is the areal density of dislocation lines, $F_1$ and $F_3$
are positive constants, $C$ is a constant of the order $a^2$ with $a$
being the lattice constant, $F_2 > 0$ at low temperatures and $F_2 < 0$
at high temperatures.  It is easy to see that Eq.~(\ref{disloc}) implies
a first order transition:
\begin{equation}
\rho = \left\{ \begin{array}{ll}
         0,                            & (T<T_m) \\
         \rho_c\exp[A(T-T_m)^\nu],     & (T \ge T_m)
            \end{array}
       \right.
\end{equation}
where $\rho_c=F_1/2F_3$ ($\approx\mu a/32\pi^2k_BT_m$, with $\mu$ being
the shear modulus, if we use the values in Ref.\ \cite{yama} for $F_1$ and
$F_3$), $A=\sqrt{-F_2^\prime/F_3\rho_c}$, and $\nu = 1/2$.  The mean
distance between dislocation lines, $\zeta = \rho^{-1/2} = \rho_c^{-1/2}
\exp[-A(T-T_m)^{\nu}/2]$, is the length scale to be compared with the
Larkin length $R_c$.  In the region where $\zeta < R_c$, the critical
current density is determined by $\zeta$:
\begin{equation}
j_c = \frac{n^{1/2}f}{V_c^{1/2}B} = \left\{ \begin{array}{ll}
 \frac{n^{1/2}f}{L_c^{1/2}R_cB}\approx\frac{n^2f^4}{C_{66}^2C_{44}r_p^3},
   & (\zeta > R_c) \\
\frac{n^{1/2}f}{L_c^{1/2}\zeta B}\approx\frac{n^{2/3}f^{4/3}\rho_c^{2/3}}
{C_{44}^{1/3}r_p^{1/3}B}\exp[\frac{2}{3}A(T-T_m)^{\nu}],  & (\zeta < R_c)
\end{array}
\right.
\label{j-3d}
\end{equation}
where in the region of $\zeta < R_c$ Eq.~(\ref{energy}) is minimized, with
$R_c$ replaced by $\zeta$, to determine $L_c$.  Since $\zeta$ has a
discontinuous jump at $T_m$ (if the transition is first order), it is
possible that $j_c$ will have a jump at the onset of the peak effect
which may occur in samples with ``very weak'' pinning (Fig.\ \ref{sketch}).
In fact, a jump in $j_c$ was observed experimentally \cite{word}.  Note
that our argument does not depend on the detailed nature of the melting
transition, e.g. first vs. second order \cite{menon}, although we have
used Eq.~(\ref{disloc}) to obtain some estimates.  The key point in our
scenario is that some other (elastically relevant) length scale enters
the system around the melting transition and it crosses the Larkin length.
In 2D, we believe this length scale is the Kosterlitz-Thouless correlation
length.  While in 3D, it is most likely that it is the length scale set
by thermally excited edge dislocations.

In conclusion, a quantitative analysis suggests that the non-local effect
of $C_{44}$ is not the cause for $j_c$ to rise with increasing temperature
at the peak effect regime.  We propose a scenario based on the melting of
the vortex lattice in the presence of weak pinning.  The onset of the peak
effect is the crossover of the two elastically relevant length scales.  We
believe this captures the basic physical picture, at least for weak enough
pinning, although ideally the analysis should be put on a more rigorous
footing.  On the other hand, if the pinning is so strong that the Larkin
length is of the order of the lattice constant, the sample should not,
according to our scenario, show the peak effect.  In this paper we have
focused only on the behavior of $j_c$.  The dynamics above $j_c$ can be
quite different in the peak regime, where a significant enhancement of
plastic flow was recently observed \cite{shobo} in 2H-NbSe$_2$.  This
indicates that the dynamically generated defects proliferate in the peak
regime as well.   We emphasize the difference between the scenario
concerning melting described in this paper and that suggested in
literature \cite{melt} in which the onset of dissipation in the presence
of a driving force or the vanishing of $j_c$ has been interpreted as the
evidence for the vortex-lattice melting transition.  The rationale behind
this suggestion is based upon the assumption that when the vortex lattice
melts, the vortex lines will flow.  This scenario may be correct if the
sample is so pure that the vortex line density is much greater than the
impurity density.
Unfortunately, this cannot be achieved even in the cleanest crystals of
2H-NbSe$_2$ where the critical current density is many orders of magnitude
smaller than that of YBCO.  Instead, the clean, high quality crystals of
2H-NbSe$_2$ always exhibit \cite{peak-shobo} a pronounced peak effect and
high-quality YBCO crystals are also found recently \cite{ling,kwok} to
show a peak effect.

{\it Note added}:  In the course of writing this manuscript, we received a
preprint by Larkin, Marchetti, and Vinokur \cite{lmv}, in which they
attribute the peak effect to the softening of $C_{66}$ just {\it below}
the melting transition.

\begin{figure}

\caption{
The temperature dependence of critical current density for a YBCO
crystal in a magnetic field, extracted from Ref. [5].
We define $T_p$ as the onset temperature at which $j_c$
starts to increase.
}
\label{data}
\end{figure}

\begin{figure}
\caption{
Schematic behavior of the two length scales $R_c$ and $\zeta$, as
functions of temperature: (a) two dimensions; (b) three dimensions with
weak pinning; and (c) three dimensions with ``very weak'' pinning.  The
temperature for the onset of the peak effect is $T_p$, where $\zeta$
becomes smaller than $R_c$.  In (c), $T_p=T_m$.  The solid part of the
lines determines the critical current density.
}
\label{sketch}
\end{figure}


\begin{references}
\bibitem{foot-kt}
We use the symbol $\zeta$, instead of the standard $\xi$, for the
Kosterlitz-Thouless correlation length to avoid confusion with the
superconducting coherence length.
\end{references}
\end{document}